\documentclass[aps,prl,twocolumn,superscriptaddress,showpacs]{revtex4}
\usepackage{bm}

\usepackage{graphicx}
\usepackage[usenames,dvipsnames]{xcolor}
\usepackage{dcolumn}

\newcolumntype{w}[1]{D{.}{.}{#1}}

\begin{document}
\preprint{Version 0.9}

\title{Phenomenological Universalities: Coherence, Supersymmetry and Growth}

\author{Marcin Molski}

\affiliation{Theoretical Chemistry Department, Adam Mickiewicz University, Umultowska 89b, 61-614 Pozna{\'n}, Poland}

\date{\today}

\begin{abstract}

The phenomenological universalities (PU) are extended to include time-depended quantum oscillatory phenomena, coherence and supersymmetry. It will be proved that this approach generates minimum uncertainty coherent states of time-dependent oscillators, which in the dissociation (classical) limit reduce to the functions describing growth (regression) of the systems evolving over time. The results obtained reveal existence of a new class of macroscopic quantum (or quasi-quantum) phenomena, which may play a vital role in coherent formation of the specific growth patterns in complex systems.

\end{abstract}

\pacs{89.75.-k, 11.10.Lm, 12.60.Jv, 03.65.Fd, 89.75.Fb}
\maketitle

The concept of PU introduced by Castorina, Delsanto, and Guiot (CDG) \cite{Castorina 2006} concerns ontologically different systems, in which  miscellaneous emerging patterns are described by the same mathematical formalism.  Universality classes are useful for their applicative relevance and facilitate the cross fertilization among various fields of research, including  physics, chemistry, biology, engineering, economics and social sciences \cite{Castorina 2006}-\cite{Delsanto 2009}. This strategy is extremely important, especially for the export of ideas, models and methods developed in one discipline to another and vice versa. The PU approach is also a useful tool for investigation of the complex systems whose dynamics is governed by nonlinear processes. Hence,  this methodology can be employed \cite{Castorina 2006} to obtain different functions of growth widely applied in actuarial mathematics, biology and medicine. In this work the research area is extended to include in the CDG scheme the quantum coherence and supersymmetry. In particular it will be proved that CDG approach can be used to generate quantum coherent states of the time-dependent Morse \cite{Morse 1929} and Wei \cite{Hua Wei 1990} oscillators, which in the dissociation (classical) limit reduce to the well-know Gompertz \cite{Gompertz 1825} and West-Brown-Enquist (WBE)-type \cite{West 2001} functions (e.g. logistic, exponential, Richards, von Bertalanffy) describing sigmoidal (S-shaped) growth of complex systems.

	According to the CDG concept, various degrees of nonlinearity appearing in the systems under consideration can be classified using the set of nonlinear equations \cite{Castorina 2006}
\begin{equation}
\frac{d\psi(q)}{dq}-x(q)\psi(q)=0,\hskip1cm \frac{dx(q)}{dq}+\Phi(x)=0.
\label{01}
\end{equation}
Here, $q=u_t t$ denotes dimensionless temporal variable, $u_t$ is a scaling factor, whereas $\Phi(x)$ stands for a generating function, which expanded into a series of $x$-variable (it slightly differs from  the original CDG formulae) \cite{Castorina 2006}
\begin{equation}
\Phi(x)=c_1(x+c_0/c_1)+c_2(x+c_0/c_1)^2+...
\label{02}
\end{equation}
produces different functions of growth $\psi(q)$ for a variety of patterns emerging in complex systems. To obtain their explicit forms a combination of Eqs. (1) is integrated to generate the growth functions \cite{Castorina 2006}
\begin{equation}
\psi(q)=\exp\left[-\int_x\frac{xdx}{\Phi(x)}+C\right]=\exp\left[\int_q x(q)dq + C\right]
\label{03}
\end{equation}
for different powers $n=1,2,...$ of the truncated series (\ref{02}). The integration constant $C$ can be calculated from a boundary condition for $x(q)$ at $q=0$. For example for $x(0)=1$, $c_0=0$, $c_1=1$, $\Phi(x)=x$ one gets the Gompertz function, whereas for $\Phi(x) = x + c_2x_2$ the allometric WBE-type function can be derived \cite{Castorina 2006}
\begin{eqnarray}
\psi(q)_G&=&\exp\left[(1-\exp(-q)\right],\nonumber\\
\psi(q)_W&=&\exp\left[1+c_2-c_2\exp\left(-q\right)\right]^{1/c_2}.
\label{04}
\end{eqnarray}
Employing this approach the PU can be classified as $U1$ $(n=1)$, $U2$ $(n=2)$ etc. with respect to the different levels of nonlinearity utilized by the complex systems during formation of the specific growth patterns. Unfortunately, the CDG approach in its original form does not take into account a very important phenomenon of regression (decay) appearing in biological, medical, demographic and economic systems. For example, in cancer biology such a situation appears under chemotherapeutic treatment of tumors subjected to cycle specific (or nonspecific) drugs causing regression of cancer \cite{Sulivan 1972}. To include this phenomenon in the CDG scheme, the first of Eqs.(\ref{01}) should be modified to the form
\begin{eqnarray}
\frac{d\psi(q)^{\dagger}}{dq}&+&x(q)\psi(q^{\dagger})=0,\nonumber \\
\psi(q)_G^{\dagger}&=&\exp\left[-(1-\exp(-q)\right],\nonumber \\
\psi(q)_W^{\dagger}&=&\exp\left[1+c_2-c_2\exp\left(-q\right)\right]^{-1/c_2},
\label{05}
\end{eqnarray}
which for $\Phi(x) = x$ and $\Phi(x) = x + c_2x^2$ produces Gompertz and WBE-type functions of regression, which decay with time. 	

Analysis of Eqs. (\ref{01}) and (\ref{05}) reveals that they can be interpreted in the framework of temporal version \cite{Molski 2010} of the space-dependent quantum supersymmetry (SUSYQM) \cite{Zhang 1990}, used among others to construct coherent states of oscillators and to obtain exact solutions of the Schr\"odinger equation for vibrating harmonic and anharmonic systems. In view of this, it is tempting to apply CDG methodology to generate the coherent states of time-dependent anharmonic oscillators and compare them with those obtained using algebraic procedure \cite{Molski 2006}. To prove that mathematical formalism of PU is a hidden form of time-dependent supersymmetry, lets differentiate growth equation (\ref{01}) once with respect to $q$-coordinate and then rearrange the derived formulae to obtain the second order differential equation in a standard eigenvalue form
\begin{eqnarray}
\frac{d^2\psi(q)}{dq^2}-\psi(q)\frac{dx(q)}{dq}-x(q)\frac{d\psi(q)}{dq}&=&\nonumber\\
\left[-\frac12\frac{d^2}{dq^2}+V(q)-P\right]\psi(q)&=& 0,
\label{06}
\end{eqnarray}
in which
\begin{equation}
V(q)-P=\frac12\left[ x(q)^2+\frac{dx(q)}{dq} \right]
\label{07}
\end{equation}
represents (with accuracy to multiplicative constant) the time-dependent version of the Riccati equation, whose spatial form is widely used in SUSYQM \cite{Zhang 1990}. The quantity $x(q)$ appearing in Eqs.(\ref{01}) and (\ref{07}) has a dual interpretation: in algebraic methods $+x(q)$ represents an anharmonic variable \cite{Cooper 1992}, whereas in SUSYQM, $- x(q)=W(q)$ stands for a superpotential \cite{Zhang 1990}, which permits construction of the supersymmetric quantal equations straightforward to analytical solution. This quantity enables also to associate via Eq.(\ref{07}) a potential energy $V(q)$ and eigenvalue $P$ with all types of functions $\psi(q)$ derived in CGD scheme. On the other hand, $\psi(q)$ in Eq. (\ref{06}) is interpreted as the solution of the differential equation (\ref{06}), whose form resembles the non-relativistic version of the quantal Feinberg equation \cite{Feinber 1967} derived by Horodecki \cite{Horodecki 1988}, which is a space-like counterpart of the time-like Schr\"odinger formula.

To prove that the CDG approach produces not only classical (macroscopic) growth functions but also quantal (microscopic) once, one may apply a linear expansion of the generating function $\Phi(x) = c_0 +c_1x$, which includes a constant term $c_0$ omitted in the CDG scheme and $c_1$ coefficient, which in the CDG approach was constrained to $1$ \cite{Castorina 2006}. Additionally, we assume that $x(0) = (1-c_0)/c_1$, which for $c_0 = 0$, $c_1 = 1$ gives the CDG initial condition $x(0) = 1$. Employing Eqs. (\ref{01}) and (\ref{03}) by integration one gets ($c_0, c_1 > 0$)
\begin{eqnarray}
x(q)&=&\frac{1}{c_1}\left[\exp(-c_1q) - c_0\right],\nonumber \\
\psi(q)&=& \exp\left\{\frac{1}{c_1^2}\left[1-\exp(-c_1q)\right] \right\}\exp\left(  -\frac{c_0}{c_1}q\right),
\label{08}
\end{eqnarray}
which by making use of correspondences $c_1^2=2x_e$, $c_0=1-x_e$ can be converted to equations
\begin{eqnarray}
x(q)&=&\left[\frac{\exp(-\sqrt{2x_e}q)-1+x_e}{\sqrt{2x_e}}\right],\nonumber\\
\psi_0(q)&=& \exp\left[\frac{1-\exp\left( -\sqrt{2x_e}q\right)}{2x_e} \right]\exp\left[ \frac{(x_e-1)q}{\sqrt{2x_e}} \right]
\label{09}
\end{eqnarray}
recently obtained by Molski \cite{Molski 2006}. Taking advantage of the Riccati equation (\ref{06}) one may derive the second order differential equation whose solution is function (\ref{09})
\begin{equation}
\left\{-\frac12\frac{d^2}{dq^2}+\frac{1}{4x_e}\left[ 1-\exp(-\sqrt{2x_e}q)\right]^2-P_0\right\}\psi_0(q)=0,
\label{10}
\end{equation}
which includes eigenvalue $P_0=1/2-x_e/4$ being a ground state ($v=0$) version of a general formulae \cite{Molski 2006} (in dimensionless unit)
$P_v=v+1/2-x_e(v+1/2)^2$, $v=0,1,2...$. It is interesting to note that Eq. (\ref{10}) under substitutions
$x_e={\hbar\omega}/{4D_e}$, $\omega=a\sqrt{{2D_e}/{mc^2}}$ and $q={a(t-t_0)}/{\sqrt{2x_e}}$ converts to the explicit form of the space-like Feinberg-Horodecki equation \cite{Horodecki 1988}
\begin{equation}
\left\{-\frac{\hbar^2}{2mc^2}\frac{d^2}{dt^2}+D_e\left[1-\exp\left[-a(t-t_0)\right]\right]^2-P_0c\right\}\psi_0(t)=0
\label{13}
\end{equation}
for the ground state of the time-dependent Morse oscillator. Here $P_0=(\hbar\omega/c)(1/2-x_e/4)$ stands for a momentum eigenvalue representing zero-point momentum of vacuum \cite{Feigel 2004, Tiggelen 2006}, $x_e$ is anharmonicity constant, $\omega$ - frequency, $D_e$ - dissociation constant, $m$ - mass, $c$ - light velocity.

Proceeding in the same manner as for the first term of Eq. (\ref{02}) one can derive $x(q)$ and $\psi(q)$ for the second order expansion of $\Phi(x) = c_1(x+ c_0/c_1)+c_2(x+ c_0/c_1)^2$ and identical as before initial condition $x(0) = (1-c_0)/c_1$
\begin{eqnarray}
x(q)&=&\frac{c_1\exp(-c_1q)}{c_1^2+c_2-c_2\exp(-c_1g)}-\frac{c_0}{c_1},\nonumber\\
&=&\frac{(sc_1/c_2)\exp[-c_1(q-g_0)]}{1-s\exp[-c_1(q-q_0)]}-\frac{c_0}{c_1},\nonumber\\
\psi(q)&=&\left\{1-s\exp[-c_1(q-q_0)]\right\}^{1/c_2}\nonumber\\
& &\left\{s\exp[-c_1(q-q_0)]\right\}^{c_0/c_1^2},\nonumber\\
q_0&=&\frac{1}{c_1}\ln\left[ \frac{2c_1^2-c_1^2c_2+2c_0c_2}{(2c_0+c_1^2)(c_1^2+c_2)}\right],\nonumber\\
s&=&\frac{c_2(2c_0+c_1^2)}{2c_1^2-c_1^2c_2+2c_0c_2}
\label{14}
\end{eqnarray}
and then one may construct the quantal Feinberg-Horodecki equation
\begin{equation}
\left\{-\frac12\frac{d^2}{dq^2}+D\left[\frac{1-\exp[-c_1(q-q_0)]}{1-s\exp[-c_1(q-q_0)]}\right]^2-P\right\}\psi(q)=0,
\label{15}
\end{equation}
for the ground state of the time-dependent Wei oscillator \cite{Hua Wei 1990} in which $q=t$, $D={(2c_0+c_1^2)^2}/{8c_1^2(1-c_2)}={mc^2D_e}/{\hbar^2}$ and  $P=D-{c_0^2}/{2c_1^2}={mc^3P_0}/{\hbar^2}$. The solution (\ref{14}) and parameters appearing in (\ref{15}) can be rewritten to the form applied by Wei \cite{Hua Wei 1990} by replacements $c_1=b$, $s=c$, $1/c_2=\rho+1/2$, $c_0/c_1^2=\rho_0$.

Continuing the search for further analogies, we find that equations of growth (\ref{01}) and regression (\ref{05}) can be specified for $\alpha=\alpha^*=0$ in the forms
\begin{eqnarray}
&\hat{A}|\alpha\rangle=\alpha\psi(q)\exp[\sqrt{2}\alpha q],
\langle\alpha|\hat{A}^{\dagger}=\alpha^*\psi(q)^{\dagger}\exp(\sqrt{2}\alpha^*q),&\nonumber\\
&\hat{A}=\frac{1}{\sqrt{2}}\left[\frac{d}{dq}-x(q)\right],\hat{A}^{\dagger}=\frac{1}{\sqrt{2}}\left[-\frac{d}{dq}-x(q)\right],&
\label{18}
\end{eqnarray}
here $[\hat{A},\hat{A}^{\dagger}]=-{dx(q)}/{dq}=\Phi(x)$,
familiar in supersymmetric theory of minimum uncertainty coherent states of space-dependent oscillators \cite{Zhang 1990}. In this formalism, $\hat{A}$ and $\hat{A}^{\dagger}$ represent annihilation and creation operators, respectively. The coherent states, which minimize the generalized position-momentum (local states) or time-energy (nonlocal states) uncertainty relations are eigenstates of the annihilation operator. They not only minimize the Heisenberg relations, but also maintain those relations in time (space) due to their temporal  (spatial) stability, hence they are called ${\it intelligent}$ coherent states \cite{Aragone 1974}. To prove that coherent states (\ref{18}) minimize the generalized time-energy uncertainty relation ($\hbar=1$)
\begin{equation}
\left[\Delta x(q)\right]^2\left(\Delta E\right)^2\geq \frac14\langle\alpha|\Phi(x)|\alpha\rangle^2,\ \Phi(x)=-i\left[x(q),\hat{E}\right],
\label{19}
\end{equation}
in which $\hat{E}=i{d}/{dq}$ is energy operator whereas $x(q)$ plays the role of a temporal anharmonic variable associated with a given type of potential, the following relationships should be derived for normalized states $\langle\alpha||\alpha\rangle=1$
\begin{eqnarray}
\langle\alpha|x(q)|\alpha\rangle&=&-\frac{1}{\sqrt{2}}\langle\alpha|\hat{A}+\hat{A}^{\dagger}|\alpha\rangle=-\frac{1}{\sqrt{2}}\left(\alpha+\alpha^* \right),\nonumber\\
\langle\alpha|\hat{E}|\alpha\rangle&=&i\frac{1}{\sqrt{2}}\langle\alpha|\hat{A}-\hat{A}^{\dagger}|\alpha\rangle=i\frac{1}{\sqrt{2}}\left(\alpha-\alpha^* \right),\nonumber\\
2\langle\alpha|x(q)^2|\alpha\rangle&=&\left(\alpha+\alpha^* \right)^2+\langle\alpha|\Phi(x)|\alpha\rangle,\nonumber\\
-2\langle\alpha|\hat{E}^2|\alpha\rangle&=&\left(\alpha-\alpha^* \right)^2-\langle\alpha|\Phi(x)|\alpha\rangle.
\label{20}
\end{eqnarray}
Having derived Eqs.(\ref{20}) we can pass to calculate the squared standard deviations
\begin{eqnarray}
\Delta x(q)^2&=&\langle\alpha|x(q)^2|\alpha\rangle-\langle\alpha|x(q)|\alpha\rangle^2=\frac12\langle\alpha|\Phi(x)|\alpha\rangle,\nonumber\\
\Delta E^2&=&\langle\alpha|\hat{E}^2|\alpha\rangle-\langle\alpha|\hat{E}|\alpha\rangle^2=\frac12\langle\alpha|\Phi(x)|\alpha\rangle,
\label{21}
\end{eqnarray}
which prove that
\begin{equation}
\left[\Delta x(q)\right]^2\left(\Delta E\right)^2=\frac14\langle\alpha|\Phi(x)|\alpha\rangle^2.
\label{22}
\end{equation}
Eq.(\ref{22}) is satisfied both for $\alpha\neq 0$ as well as $\alpha=0$ and an arbitrary form of generating function $\Phi(x)$. Those facts indicate that $\psi(q)$ in CDG approach can be interpreted as a minimum uncertainty coherent state of the time-dependent oscillators characterized by anharmonic variable $x(q)$. It is noteworthy that  this interpretation remains independent of the type of generating function $\Phi(x)$, hence it can be applied both to micro- and macroscopic systems, characterized by $c_0\neq 0$ and $c_0=0$, respectively. In particular, using the CDG approach one may construct the coherent states of the time-dependent Morse oscillator, which for $c_0=0$ and $c_1=1$ convert to the Gompertzian coherent states of growth (regression) first time derived by Molski and Konarski \cite{Molski 2003}
\begin{eqnarray}
&\frac{1}{\sqrt{2}}\left\{ \frac{d}{dq} -\frac{1}{c_1}\left[\exp(-c_1q)-c_0\right]\right\}|\alpha\rangle=&\nonumber\\
&\alpha\exp\left\{\frac{1}{c_1^2}\left[1-\exp(-c_1q)\right]\right\}\exp\left[-\frac{c_0 q}{c_1}\right]\exp(\sqrt{2}\alpha q)&\nonumber\\
&\stackrel{c_0=0, c_1=1}\longrightarrow\frac{1}{\sqrt{2}}\left\{ \frac{d}{dq} -\left[\exp(-q)\right]\right\}|\alpha\rangle=&\nonumber\\
&\alpha\exp\left[1-\exp(-q)\right]\exp(\sqrt{2}\alpha q)\stackrel{\alpha=0}\longrightarrow& \nonumber\\
&\left[ \frac{d}{dq} -\exp(-q)\right]\exp\left[1-\exp(-q)\right]=0,&
\label{23}
\end{eqnarray}
\begin{eqnarray}
&\langle\alpha|\frac{1}{\sqrt{2}}\left\{ -\frac{d}{dq} -\frac{1}{c_1}\left[\exp(-c_1q)-c_0\right]\right\}=&\nonumber\\
&\alpha^*\exp\left\{-\frac{1}{c_1^2}\left[1-\exp(-c_1q)\right]\right\}\exp\left[\frac{c_0 q}{c_1}\right]\exp(\sqrt{2}\alpha^* q) &\nonumber\\
&\stackrel{c_0=0, c_1=1}\longrightarrow\langle\alpha|\frac{1}{\sqrt{2}}\left\{ -\frac{d}{dq} -\left[\exp(-q)\right]\right\}=&\nonumber\\
&\alpha^*\exp\left\{-\left[1-\exp(-q)\right]\right\}\exp(\sqrt{2}\alpha^* q)\stackrel{\alpha^*=0}\longrightarrow& \nonumber\\
&\left[ -\frac{d}{dq} -\exp(-q)\right]\exp\left\{-\left[1-\exp(-q)\right]\right\}=0.&
\label{24}
\end{eqnarray}
In a similar manner, one may construct the coherent states of time-dependent Wei oscillator, which in the dissociation (classical) limit convert to the coherent WBE-type function of growth
\begin{eqnarray}
&\frac{1}{\sqrt{2}}\left\{ \frac{d}{dq}-\frac{(sc_1/c_2)\exp\left[-c_1(q-q_0)\right]}{1-s\exp\left[-c_1(q-q_0)\right]}+\frac{c_0}{c_1}\right\}|\alpha\rangle=&\nonumber\\
&\alpha\left\{1-s\exp\left[-c_1(q-q_0)\right]\right\}^{1/c_2}\left\{s\exp\left[-c_1(q-q_0)\right]\right\}^{c_0/c_1^2}&\nonumber\\
&\exp(\sqrt{2}\alpha q)&\nonumber\\
&\stackrel{c_0=0, c_1=1}\longrightarrow  \frac{1}{\sqrt{2}}\left\{ \frac{d}{dq}-\frac{(s^{\prime}/c_2)\exp\left[-(q-q^{\prime}_0)\right]}{1-s^{\prime}\exp\left[-(q-q^{\prime}_0)\right]}\right\}|\alpha\rangle=&\nonumber\\
&\alpha\left\{1-s^{\prime}\exp\left[-(q-q^{\prime}_0)\right]\right\}^{1/c_2}\exp(\sqrt{2}\alpha q)\stackrel{\alpha=0}\longrightarrow &\nonumber\\
&\left\{ \frac{d}{dq}-\frac{(s^{\prime}/c_2)\exp\left[-(q-q^{\prime}_0)\right]}{1-s^{\prime}\exp\left[-(q-q^{\prime}_0)\right]}\right\}\left\{1-s^{\prime}
\exp\left[-(q-q^{\prime}_0)\right]\right\}^{1/c_2}=&\nonumber\\
&\left[ \frac{d}{dq}-\frac{\exp\left(-q\right)}{1+c_2-c_2\exp\left(-q\right)}\right]\left[1+c_2
-c_2\exp\left(-q\right)\right]^{1/c_2}=0.&
\label{25}
\end{eqnarray}
Here $q^{\prime}_0=\ln\left[(2-c_2)/(1+c_2)\right]$ and $s^{\prime}=c_2/(2-c_2)$. Analogically the WBE states of regression can be derived from quantal solutions of the creation equation (\ref{18}). The results obtained indicate that the concept of PU originally applied only to macroscopic complex systems can be extended to include quantum phenomena such as coherence and supersymmetry playing a vital role on the microscopic level. In connection presented a  micro-macro conversion  is accomplished by $c_0\rightarrow 0$, which transforms quantum equations into classical ones. Only one exception is uncertainty relation (\ref{19}), which is satisfied both for micro- and macroscopic functions $\psi(q)$ generated for an arbitrary form of $\Phi(x)$. This fact has very important interpretative implications. The time-like coherent states, which minimize the position-momentum uncertainty relation evolve coherently in time being localized on the classical space-trajectory \cite{Zhang 1990}. On the contrary, the space-like coherent states which minimize the time-energy uncertainty relation evolve along localized (classical) time-trajectory being coherent in all points of space \cite{Molski 2003, Molski 2006}. Such states assumed to be coherent at an arbitrary point of space remain coherent in all points of space. We conclude that the spatial coherence is an immanent feature of all systems whose growth (decay) is described by functions derived in the CDG scheme independently of their quantal or classical nature. Although the notions of coherence and supersymmetry are usually attributed to microscopic systems, the correspondence principle introduced by Niels Bohr \cite{Rosenfeld 1976} allows for the physical characteristics of quantum systems to be maintained also in classical regime. According to this concept, the quantum theory of  micro-objects passes asymptotically into the classical one when the quantum numbers characterizing the micro-system attain extremely high values or we can neglect the Planck's constant. In this way one may derive e.g. from quantal Planck's black-body radiation formula the classical Rayleigh-Jeans law describing the spectral radiance of electromagnetic waves.  Both models describe the same phenomenon but employ diverse (quantum vs classical) formalisms and are valid for different wavelength ranges of emitted radiation. Identical situation appears in the case of quantal oscillatory phenomena which in the classical limit possess the same characteristics as their quantum counterparts. The first- and second-order growth equations obtained in this way do not contain mass nor Planck's constant \cite{Molski 2003, Molski 2006}, therefore according to the correspondence principle, they represent classical equations of coherent growth (regression). It is straightforward to demonstrate that for $c_0=0$ quantal Eqs. (\ref{10}),(\ref{15}),(\ref{23}),(\ref{24}),(\ref{25}) convert to their classical counterparts characterized by the dissociation condition $P=D$. We conclude that the macroscopic Gompertz and WBE-type functions have identical forms as microscopic ground state solutions of the Feinberg-Horodecki equation for time-dependent Morse and Wei oscillators in the dissociation state. In this limit the direction of temporal growth (regression) is consistent with the arrow of time - it is not of the oscillatory type as predicted for hypothetical bound states of time-dependent oscillators. The extension of the PU strategy presented in this work permits including in the CDG classification scheme the coherence and supersymmetry persisting both in micro- and macro domains. Hence, the results obtained reveal existence of a new class (according to the Leggett classification \cite{Leggett 1987}) of macroscopic quantum (or quasi-quantum) phenomena, which may play a vital role in coherent formation of the specific growth patterns in complex systems. The method presented can be employed also to the space-dependent phenomena using $q=u_rr$ spatial variable in which $u_r$ is a scaling factor. In this way one may generate in the CDG scheme the coherent states of the space-dependent Morse and Wei oscillators, which minimize the position-momentum uncertainty relation \cite{Molski 2009} and in dissociation limit $c_0\rightarrow 0$ or, equivalently $E\rightarrow D$, reduce to the space-dependent sigmoidal Gompertz and WBE-like functions widely applied in a range of fields including e.g.  probability theory and statistics where are used to describe cumulative distribution of entities characterized by different spatial sizes \cite{Easton 2005}.

\end{document}